\documentclass[aps,twocolumn,pra,superscriptaddress,amsmath,showpacs,tightenlines]{revtex4-1}
\usepackage{amssymb}
\usepackage{amsmath}
\usepackage{esint}
\usepackage{graphicx}
\usepackage{subfigure}
\usepackage{natbib}
\usepackage{epsfig}
\usepackage{amsfonts}
\usepackage{mathrsfs}
\usepackage{pifont}
\usepackage{xcolor}
\usepackage[toc,page,title,titletoc,header]{appendix}
\begin{document}

\title{Rabi oscillation and fractional population via the bound states in the continuum in a giant atom waveguide QED setup}

\author{Hongwei Yu}
\affiliation{Center for Quantum Sciences and School of Physics, Northeast Normal University, Changchun 130024, China}
\author{Xiaojun Zhang}
\affiliation{Center for Quantum Sciences and School of Physics, Northeast Normal University, Changchun 130024, China}
\author{Zhihai Wang}
\email{wangzh761@nenu.edu.cn}
\affiliation{Center for Quantum Sciences and School of Physics, Northeast Normal University, Changchun 130024, China}
\author{Jin Wang}
\email{jin.wang.1@stonybrook.edu}
\affiliation{Department of Chemistry and of Physics and  Astronomy, Stony Brook University, Stony Brook, NY 11794-3400, USA}

\begin{abstract}
We study the dynamics of two giant atoms interacting with a coupled resonator waveguide (CRW) beyond the Markovian approximation. The distinct atomic configurations determine the number of bound states in the continuum (BIC), leading to different dynamical behaviors. Our results show that when the system supports two BICs, Rabi oscillations dominate the dynamics, whereas fractional population dynamics emerge in the presence of a single BIC. The connection between these dynamics and the existence of BICs is further verified by analyzing the photonic distribution in the CRW during time evolution. These findings challenge the conventional notion that the environment always induces dissipation and decoherence. Instead, the bound states in the CRW-emitters coupled system can suppress complete dissipation of the emitters. This work offers an effective approach for controlling dissipative dynamics in open quantum systems.

 \end{abstract}

\maketitle
\section{introduction}

One of the core areas of interest in waveguide quantum electrodynamics is the exploration of interactions between atoms and photons, which forms the foundation for many quantum technologies~\cite{AD2015,CD2019}. Traditionally, it is assumed that the size of atoms is much smaller than the wavelength of light, allowing atoms to be modeled as point dipoles~\cite{DL2003}. In 2014, however, superconducting artificial atoms~\cite{JK2007} were successfully coupled to surface acoustic waves, where the atom size becomes comparable to the wavelength of the surface acoustic waves~\cite{SD1986,DM2007}. This marked the emergence of the concept of the ``giant atom" as the atom could no longer be approximated as a point~\cite{MV2014}.
Since then, the giant atom model has been realized in various platforms, including transmon qubits~\cite{AF2018,BK2020,GA2019,BK2020,AM2021} and magnon systems~\cite{ZW2022}. The central feature of giant atoms is the breaking of the dipole approximation~\cite{DF2008}, induced by the nonlocal coupling between the atom and the waveguide~\cite{AF2021,WZ2020}.

The waveguide provides a channel for photonic propagation and serves as a data bus to mediate interactions between remote emitters~\cite{XG2017,DR2017,RH19701,RH19702,KL2013,HZ2013}. In the newly developed giant atom setup, photon propagation between multiple atom-waveguide coupling points generates intriguing interference effects, such as decoherence-free interactions~\cite{AF2018,AC2020} and non-Markovian oscillations~\cite{LG2017,GA2019,CA2021,LG2020,SG2020}.
Previous studies often considered giant atoms coupled to continuous waveguides~\cite{XL2022,LD2023,SLF2021,YC2022}, which support a linear dispersion relation and an infinite energy spectrum. Leveraging advancements in photonic crystal~\cite{YL2017,NM2019,PM2019} and superconducting quantum circuit technologies~\cite{GA2019,BK2020,AM2021}, the coupled resonator waveguide (CRW) has been realized. This structured environment features both a continuous band and a band gap, enabling a richer range of dynamics. As a result, the interactions between giant atoms mediated by the CRW have recently become a topic of significant interest.


In this paper, we consider two giant atoms coupled to a coupled resonator waveguide (CRW) in a braided configuration. We find that the system dynamics are closely related to the presence of bound states in the continuum (BIC). The concept of BIC dates back to 1975~\cite{FH1975}, and there are two primary mechanisms for its formation: (1) the zero density of states at certain frequencies in the environment, and (2) decoupling between the system and the environment due to interference effects~\cite{DC2008,MI2012,GC2019,QQ2023}. Although BICs lie within the continuous spectrum in the frequency domain, they exhibit bound-state-like localization in real space. Due to the existence of BICs in our system, non-Markovian effects~\cite{LG2017} dominate the system dynamics. On the one hand, we analyze the energy spectrum of the giant atom-CRW coupled system, numerically identifying the BICs of the system. On the other hand, we derive the dynamical equations for the two giant atoms by tracing out the degrees of freedom of the CRW, going beyond the Markovian approximation. This approach allows us to predict the reduced dynamics of the quantum open system, i.e., the two giant atoms in our setup. As a result, we establish a consistent connection between the global properties of the energy spectrum and the reduced dynamics of the giant atoms.

Generally, it is widely understood that the environment drives an open quantum system into the vacuum state without any excitation as the evolution time approaches infinity. A notable exception is the subradiance process, associated with dark states that are decoupled from the environment~\cite{wang2015,NS2018}. In our system, however, we find that the reduced dynamics of the giant atoms exhibit behaviors beyond subradiance, determined by the number of bound states in the continuum (BIC). When the system supports two BICs, the two giant atoms exhibit persistent Rabi oscillations~\cite{SL2021,KH2023,CW2016,MF2017,KK2019,KKK2019} without dissipating into the waveguide. In contrast, with a single BIC, we observe fractional population dynamics~\cite{JT2022}, where the giant atoms maintain finite excitation even as the evolution time approaches infinity. Notably, similar behavior is also observed in non-Markovian processes associated with bound states outside the continuum~\cite{ELi1987,Wang1990,An2010,An2013}. Furthermore, our non-Markovian treatment enables us to analyze the dynamical behavior of the photonic distribution in the waveguide, providing additional evidence for the connection between the system dynamics and the presence of BICs.

The rest of the paper is organized as follows.
In Sec.~\ref{modelsection}, we introduce the model and analyze the connection between the dynamical equation and the BIC of the system. In Sec.~\ref{dynamicssection}, we present two typical dynamical behaviors of the system: Rabi oscillations and fractional population dynamics, which occur when the system is characterized by two and one BIC, respectively.
In Sec.~\ref{consection}, we provide a brief conclusion. Detailed derivations of the bound states and the dynamical equations are presented in the Appendices.

\section{Model}
\label{modelsection}

\subsection{Hamiltonian}
\begin{figure}
\begin{centering}
\includegraphics[width=1.0\columnwidth]{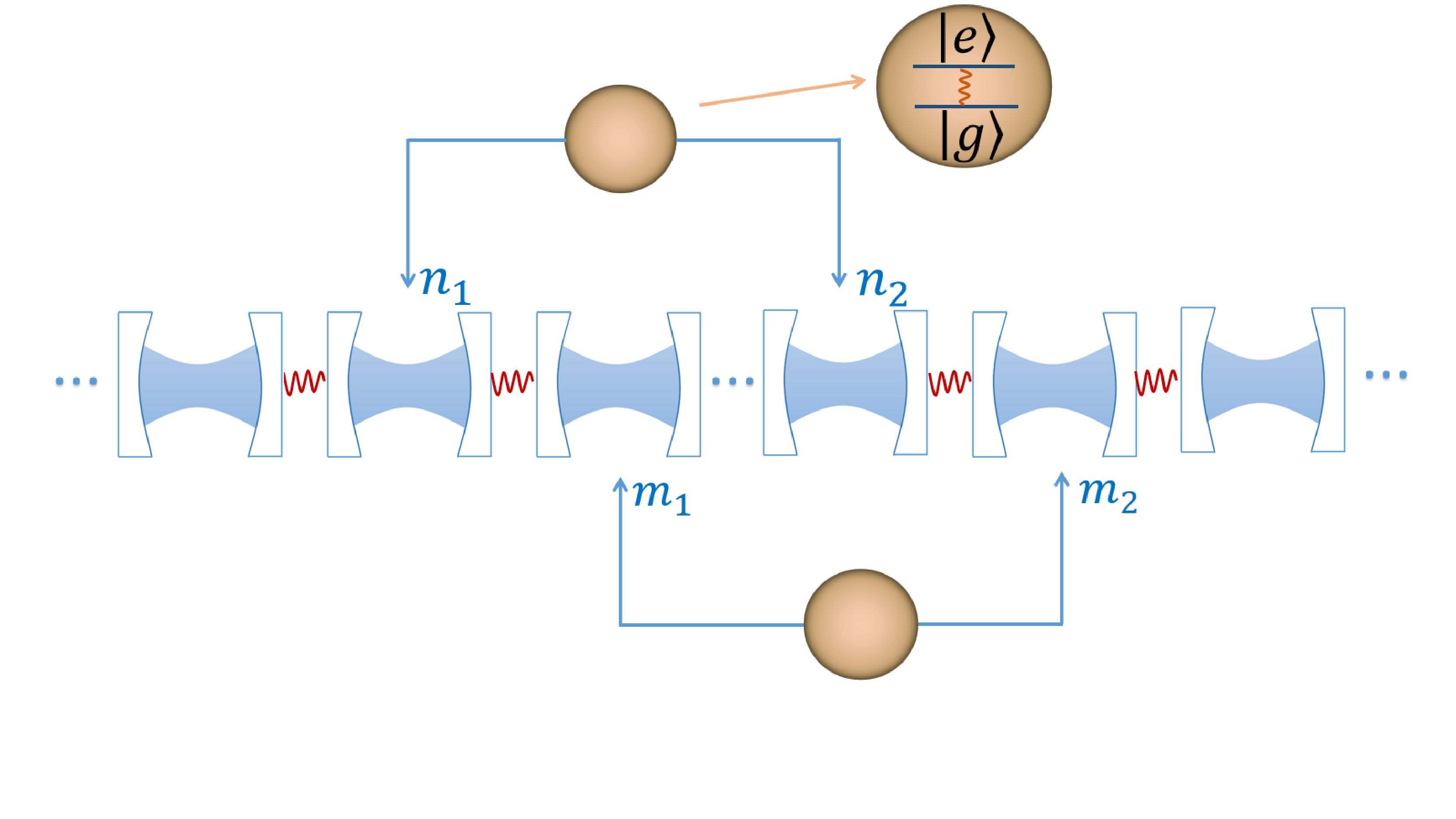}
\end{centering}
\caption{Schematic diagram of the coupling between two braided giant atoms and the coupled resonator waveguide (CRW). In the rest of the paper, we refer to the left (right) atom, which couples to the CRW via sites $n_1$ and $n_2$ ($m_1$ and $m_2$), as the first (second) atom.}
\label{setup}
\end{figure}

As schematically shown in Fig.~\ref{setup}, the system under consideration consists of two braided giant atoms coupled to a one-dimensional coupled resonator waveguide (CRW) via two separate sites, respectively~\cite{AF2018}. The Hamiltonian of the system is divided into three parts, $H = H_{a} + H_{c} + H_{I}$, where (hereafter, we set $\hbar = 1$):
\begin{eqnarray}
H_{a} &=& \Omega_{1}\left|e\right\rangle_{1}\left\langle e\right| + \Omega_{2}\left|e\right\rangle_{2}\left\langle e\right|, \\
H_{c} &=& \omega_{c}\sum_{j}a_{j}^{\dagger}a_{j}
- \xi\sum_{j}\left(a_{j+1}^{\dagger}a_{j} + a_{j}^{\dagger}a_{j+1}\right), \\
H_{I} &=& g_{1}\left(a_{n_{1}}^{\dagger}\sigma_{1}^{-} + a_{n_{2}}^{\dagger}\sigma_{1}^{-} + {\rm H.c.}\right) \nonumber \\
& & + g_{2}\left(a_{m_{1}}^{\dagger}\sigma_{2}^{-} + a_{m_{2}}^{\dagger}\sigma_{2}^{-} + {\rm H.c.}\right).
\end{eqnarray}

Here, $H_a$ describes the Hamiltonian of the two two-level giant atoms, where $\Omega_{1}$ ($\Omega_{2}$) is the transition frequency of the first (second) giant atom between the excited state $|e\rangle$ and the ground state $|g\rangle$. $H_c$ represents the Hamiltonian of the CRW~\cite{SL2008}, where $\omega_{c}$ is the bare frequency of each identical resonator in the waveguide, $\xi$ is the photonic hopping strength between adjacent resonators, and $a_{j}^{\dagger}$ ($a_{j}$) are the creation (annihilation) operators in the $j$th resonator.

$H_I$ describes the interaction between the giant atoms and the waveguide, where $g_{1}$ ($g_{2}$) is the coupling strength between the first (second) giant atom and the waveguide. Notably, the first (second) giant atom is coupled only to the $n_{1}$th and $n_{2}$th ($m_{1}$th and $m_{2}$th) resonators in the waveguide.

Considering the number of resonators to be infinite, i.e., $N_c \rightarrow \infty$, we introduce the Fourier transform $a_{j} = \frac{1}{\sqrt{N_c}}\sum_{k}e^{-ikj}b_{k}$~\cite{SL2020} to rewrite the Hamiltonian of the waveguide $H_c$ in momentum space as:
\begin{equation}
H_{c} = \sum_{k}\omega_{k}b_{k}^{\dagger}b_{k},
\end{equation}
where the dispersion relation of the waveguide is $\omega_{k} = \omega_{c} - 2\xi\cos k$ with $k \in \left[-\pi, \pi\right)$. Here, the lattice constant is set to unity, making the wave vector $k$ dimensionless. The CRW thus provides a structured environment for the atoms, with an energy band centered at $\omega_c$ and a width of $4\xi$.

In momentum space, the total Hamiltonian becomes:
\begin{eqnarray}
H &=& \Omega_{1}\left|e\right\rangle_{1}\left\langle e\right| + \Omega_{2}\left|e\right\rangle_{2}\left\langle e\right| + \sum_{k}\omega_{k}b_{k}^{\dagger}b_{k} \nonumber \\
& & + \frac{g_{1}}{\sqrt{N_c}}\sum_{k}\left[\left(e^{ikn_{1}} + e^{ikn_{2}}\right)b_{k}^{\dagger}\sigma_{1}^{-} + {\rm H.c.}\right] \nonumber \\
& & + \frac{g_{2}}{\sqrt{N_c}}\sum_{k}\left[\left(e^{ikm_{1}} + e^{ikm_{2}}\right)b_{k}^{\dagger}\sigma_{2}^{-} + {\rm H.c.}\right].
\label{ksplaceH}
\end{eqnarray}

\subsection{Bound states}
As illustrated in the previous subsection, the waveguide supports a photonic continuum band. The two-site coupling of the giant atom does not alter the extended nature of these continuum levels but introduces two additional types of bound states.

On the one hand, the giant atom breaks the translational symmetry of the waveguide, leading to the formation of a bound state outside the continuum (BOC). Such BOCs have been identified in various open systems with structured environments, such as photonic crystals and atom-CRW coupled systems~\cite{Wang1990,Zhou2008,Zhou2013}. On the other hand, under specific parameter regimes, interference between the two coupling sites decouples the giant atom from the waveguide, leading to the BIC. To clarify this mechanism, we refer to our previous work~\cite{Yu2021,XJ2023}, which demonstrates that a single BIC can exist when only one giant atom is coupled to the CRW.

For instance, when $g_2=0$, the Hamiltonian in momentum space for a single giant atom system is:
\begin{equation}
H' = \Omega_{1}\left|e\right\rangle_{1}\left\langle e\right| + \sum_{k}\omega_{k}b_{k}^{\dagger}b_{k} + \sum_{k}\left[g_kb_{k}^{\dagger}\sigma_{1}^{-} + {\rm H.c.}\right],
\end{equation}
where $g_k = \frac{g_1}{\sqrt{N_c}}[\exp(ikn_1) + \exp(ikn_2)]$. Since the dispersion relation of the CRW satisfies $\omega_k = \omega_c - 2\xi\cos k$, the giant atom resonates with the photonic mode at $k = \pi/2$ when $\Omega_1 = \omega_c$. In this case, $g_k=0$ as long as the size of the giant atom satisfies $N=4m+2$, where $m=1,2,\dots$~\cite{XJ2023}. This implies that the giant atom decouples from the resonant photonic mode, resulting in a BIC.

In the current setup with two braided giant atoms, we consider two types of giant atoms by setting $\Omega_1 = \Omega_2 = \omega_c$. For the first type, where $N=6$, a single BIC exists in the single giant atom system. For the second type, where $N=8$, no BIC exists in the single giant atom system.

Assuming the single excitation eigenstate of the system $\left|E\right\rangle$ with eigenenergy $E$ can be expressed as:
\begin{equation}
\left|E\right\rangle = \left(A_{1}\sigma_{1}^{+} + A_{2}\sigma_{2}^{+} + \sum_{k}B_{k}b_{k}^{\dagger}\right)\left|G\right\rangle,
\end{equation}
where $|G\rangle$ represents the state where both giant atoms are in their ground states and all resonators are in their vacuum states. Here, $A_{1}$, $A_{2}$, and $B_{k}$ are the amplitudes of the first giant atom, the second giant atom, and the $k$th mode of the waveguide being excited, respectively.

The eigenvalue equation $H\left|E\right\rangle = E\left|E\right\rangle$ leads to:
\begin{eqnarray}
\Omega_{1}A_{1} + \frac{g_{1}}{\sqrt{N_{c}}}\sum_{k}\left(e^{-ikn_{1}} + e^{-ikn_{2}}\right)B_{k} &=& EA_{1}, \label{a1} \\
\Omega_{2}A_{2} + \frac{g_{2}}{\sqrt{N_{c}}}\sum_{k}\left(e^{-ikm_{1}} + e^{-ikm_{2}}\right)B_{k} &=& EA_{2}, \label{a2}
\end{eqnarray}
and
\begin{equation}
B_{k} = \frac{g_{1}}{\sqrt{N_{c}}}\frac{A_{1}\left(e^{ikn_{1}} + e^{ikn_{2}}\right)}{E-\omega_{k}} + \frac{g_{2}}{\sqrt{N_{c}}}\frac{A_{2}\left(e^{ikm_{1}} + e^{ikm_{2}}\right)}{E-\omega_{k}}. \label{bk}
\end{equation}

Substituting $B_{k}$ into Eqs.~(\ref{a1}) and (\ref{a2}) and eliminating $B_{k}$, we obtain:
\begin{eqnarray}
&& A_{1}\left[E - \Omega_{1} - \frac{g_{1}^{2}}{N_{c}}\sum_{k}\frac{2 + \cos\left[k\left(n_{2}-n_{1}\right)\right]}{E-\omega_{k}}\right] \nonumber\\
&& = A_{2}\frac{g_{1}g_{2}}{N_{c}}\sum_{k}\frac{\sum_{j,j'=1}^{2}e^{ik\left(m_{j}-n_{j'}\right)}}{E-\omega_{k}}, \label{a11}
\end{eqnarray}
\begin{eqnarray}
&& A_{2}\left[E - \Omega_{2} - \frac{g_{2}^{2}}{N_{c}}\sum_{k}\frac{2 + \cos\left[k\left(m_{2}-m_{1}\right)\right]}{E-\omega_{k}}\right] \nonumber\\
&& = A_{1}\frac{g_{1}g_{2}}{N_{c}}\sum_{k}\frac{\sum_{j,j'=1}^{2}e^{ik\left(n_{j}-m_{j'}\right)}}{E-\omega_{k}}. \label{a22}
\end{eqnarray}

Under the symmetric and resonant condition $g_{1}=g_{2}=g$, $\Omega_{1}=\Omega_{2}=\Omega$, and $m_{2}-m_{1}=n_{2}-n_{1}=N$, the transcendental equation for the eigenenergy is derived as (details are in Appendix A):
\begin{equation}
E = \Omega - \frac{g^{2}}{\xi}\left[\frac{2 + 2\chi^{N} \pm \sum_{j,j'=1}^{2}\chi^{n_{j}-m_{j'}}}{\chi - \chi^{*}}\right], \label{EBIC}
\end{equation}
where $\chi = \exp\left[-i\arccos\left(\frac{E-\omega_{c}}{2\xi}\right)\right]$.

\subsection{Dynamical equation}

In what follows, we investigate the dynamical equations of the system in the single-excitation subspace, where the time-dependent wave function can be expressed as:
\begin{equation}
\label{wavet}
\left|\psi\left(t\right)\right\rangle =\left[\alpha_{1}\left(t\right)\sigma_1^+ +\alpha_{2}\left(t\right)\sigma_2^+  +\sum_{k}\beta_{k}\left(t\right)b_k^{\dagger}\right]\left|G\right\rangle,
\end{equation}
where $|G\rangle$ represents the state in which both giant atoms are in their ground states and all resonators are in their vacuum states. The coefficients $\alpha_{1}\left(t\right)$, $\alpha_{2}\left(t\right)$, and $\beta_{k}\left(t\right)$ are the amplitudes for finding an excitation in the first giant atom, the second giant atom, and the $k$th mode in the waveguide, respectively.

After detailed calculations under the Weisskopf-Wigner approximation but beyond the Markovian approximation (see Appendix~\ref{A1} for details), the dynamics of the two giant atoms are governed by:
\begin{equation}
i \frac{d\vec{\alpha}(t)}{dt} = M(t)\vec{\alpha}(t),\label{dyn1}
\end{equation}
where $\vec{\alpha}(t) = \left(\alpha_1(t), \alpha_2(t)\right)^T$, and
\begin{equation}
M(t) = \left(\begin{array}{cc}
\mathcal{A}_{1}\left(t\right) & \mathcal{B}\left(t\right) \\
\mathcal{B}\left(t\right) & \mathcal{A}_{2}\left(t\right)
\end{array}\right),
\end{equation}
with
\begin{equation}
\mathcal{B}(t) = -i g_1 g_2 \int_0^t d\tau e^{-i\omega_{c}\tau}\sum_{j,j'=1}^2 i^{n_j-m_{j'}}J_{n_j-m_{j'}}(2\xi\tau),
\end{equation}
representing the waveguide-induced interaction between the giant atoms. Here, $J_{m}$ is the $m$th order Bessel function.

The diagonal terms in the matrix, $\mathcal{A}_i(t)$, represent the self-energy of the $i$th giant atom:
\begin{equation}
\mathcal{A}_i(t) = \Omega_{i} - 2i g_{i}^{2}\int_{0}^{t}d\tau e^{-i\omega_{c}\tau}\left\{ J_{0}(2\xi\tau) + i^{N_i}J_{N_i}(2\xi\tau)\right\},
\label{Ai}
\end{equation}
where $N_1 = n_2 - n_1$ and $N_2 = m_2 - m_1$ are the sizes of the first and second giant atoms, respectively. In Eq.~(\ref{Ai}), the first term represents the free evolution of the giant atom, while the second term accounts for the backaction of the waveguide on the atom.

As shown in the Appendix, the Weisskopf-Wigner approximation also allows us to derive the photonic dynamics in real space as:
\begin{eqnarray}
\beta_{j} &=& \frac{1}{\sqrt{N_{c}}}\sum_{k}e^{-ikj}\beta_{k} \nonumber \\
&=& -i g_{1} \int_{0}^{t}d\tau \alpha_{1}\left(t-\tau\right)F_{j}\left(\tau\right) \nonumber \\
& & -i g_{2} \int_{0}^{t}d\tau \alpha_{2}\left(t-\tau\right)G_{j}\left(\tau\right),
\label{photon}
\end{eqnarray}
where
\begin{eqnarray}
F_{j}\left(\tau\right) &=& e^{-i\omega_{c}\tau}\sum_{q=1}^{2} i^{\left(j-n_{q}\right)}J_{\left(j-n_{q}\right)}(2\xi\tau), \\
G_{j}\left(\tau\right) &=& e^{-i\omega_{c}\tau}\sum_{q=1}^{2} i^{\left(j-m_{q}\right)}J_{\left(j-m_{q}\right)}(2\xi\tau).
\end{eqnarray}

\subsection{Relation between BIC and dynamical equations}

In Ref.~\cite{SL2021}, the existence condition of BIC(s) in a two-giant-atom setup was investigated under the Markovian approximation, where the semi-negative matrix $M$ is time-independent. The author claimed that the number of BICs in the system equals the number of zero eigenvalues of $M$. The underlying physics can be explained as follows: since the system cannot continuously expand, the imaginary parts of the eigenvalues of $M$ must be non-positive if $M$ is time-independent. In this context, {if the imaginary part of an eigenvalue is zero,} it implies that the system does not dissipate into the environment (i.e., the CRW in our case). Thus, the system is decoupled from the waveguide, forming a BIC, similar to the case of a single giant atom~\cite{XJ2023}.

In our Weisskopf-Wigner approximation treatment, the matrix $M$ is time-dependent, but the relation between the BIC and the eigenvalues of $M$ still holds. In what follows, we solve the transcendental equation Eq.~(\ref{EBIC}) to determine the number and energy values of the BICs and connect them to the time-dependent eigenvalues $\lambda(t)$ of the matrix $M$.

\begin{figure}
\begin{centering}
\includegraphics[width=1.0\columnwidth]{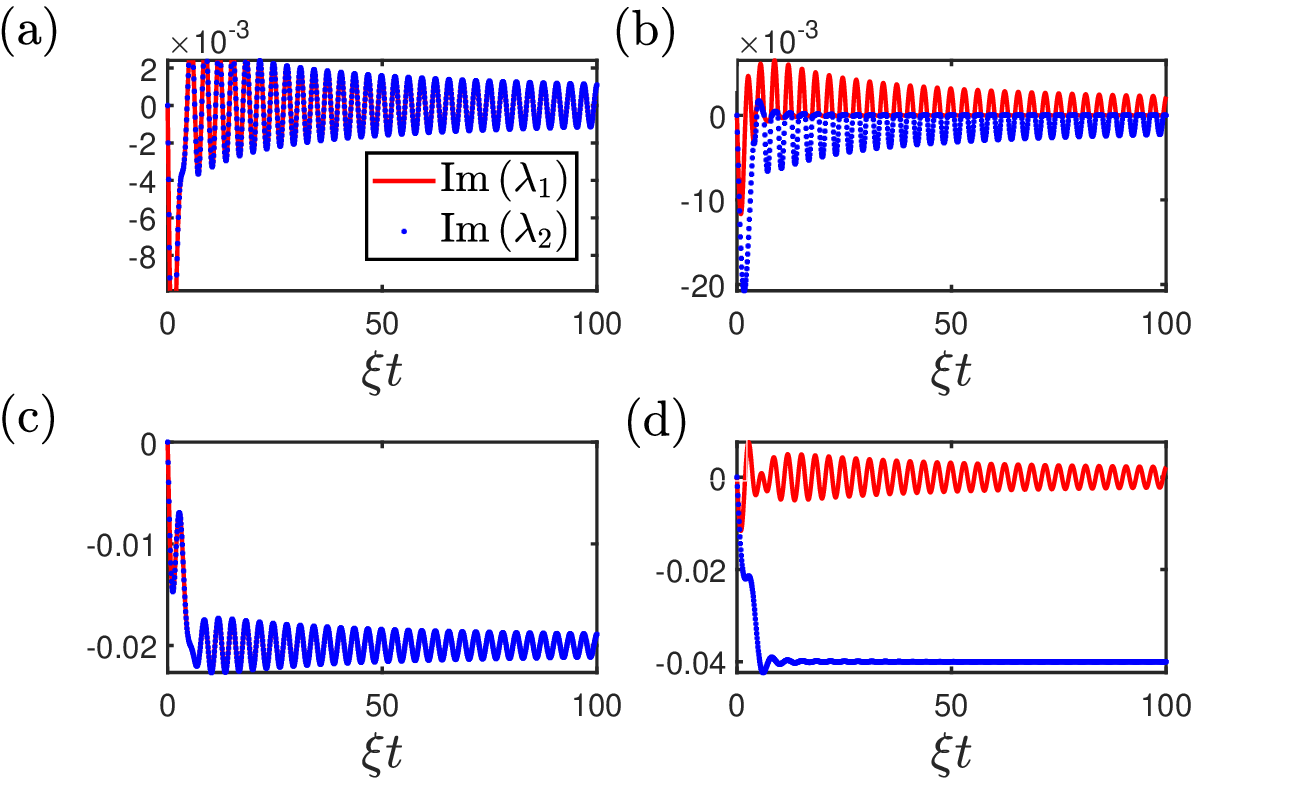}
\par\end{centering}
\caption{The imaginary parts of the eigenvalues of the matrix $M$ as functions of the evolution time.
(a) $N_1 = N_2 = 6, \Delta = 3$.
(b) $N_1 = N_2 = 6, \Delta = 2$.
(c) $N_1 = N_2 = 8, \Delta = 3$.
(d) $N_1 = N_2 = 8, \Delta = 2$.
The parameters are set as $\Omega_{1} = \Omega_{2} = \omega_{c} = 0$, and $g_{1} = g_{2} = 0.1\xi$.
}
\label{eigenvalue}
\end{figure}

The results for the solutions of Eq.~(\ref{EBIC}) are summarized in Table~I, while the corresponding eigenvalues $\lambda(t)$ as functions of evolution time are illustrated in Fig.~\ref{eigenvalue}. For $N_1=N_2=6$, as shown in Figs.~\ref{eigenvalue}(a) and (b), we observe that $\mathrm{Im}\left(\lambda_{1}\right) = \mathrm{Im}\left(\lambda_{2}\right) \approx 0$, indicating that the system has two BICs, consistent with the first two rows in Table~I. This result suggests that when a single giant atom can generate a BIC, combining two giant atoms will produce two BICs, regardless of the relative distance $\Delta$ between them.

For $N_1 = N_2 = 8$, where a single giant atom setup does not support a BIC, the presence of a BIC in the two braided giant atoms setup depends on their relative distance, as shown in the last two rows of Table~I. In the setup with $\Delta=3$, as depicted in Fig.~\ref{eigenvalue}(c), both $\mathrm{Im}\left(\lambda_{1}\right)$ and $\mathrm{Im}\left(\lambda_{2}\right)$ are negative, indicating the absence of a BIC. However, for $\Delta=2$, the result in Fig.~\ref{eigenvalue}(d) shows $\mathrm{Im}\left(\lambda_{1}\right) \approx 0$ and $\mathrm{Im}\left(\lambda_{2}\right) < 0$, implying that there is only one BIC in the system.

\begin{table}
\caption{The solutions of the transcendental equation Eq.~(\ref{EBIC}) in the regime $-2\xi < E < 2\xi$. The parameters are set as $\Omega_1 = \Omega_2 = \omega_c = 0$, and $g_1 = g_2 = 0.1\xi$.}
\begin{tabular}{|c|c|c|}
\hline
$N_{1}=N_{2}$ & $\Delta=m_{1}-n_{1}$ & $E\:(-2\xi<E<2\xi)$\tabularnewline
\hline
6 & {1,\,3,\,5} & {$E_{1}=0.0097\xi;E_{2}=-0.0097\xi$} \tabularnewline
\hline
6 & {2,\,4} & {$E_{1}=E_{2}=0$} \tabularnewline
\hline
8 & {1,\,3,\,5,\,7} & {No BIC} \tabularnewline
\hline
8 & {2,\,4,\,6} & {$E=0$} \tabularnewline
\hline
\end{tabular}
\end{table}

\section{Dynamics}
\label{dynamicssection}

Based on the results in Table I, we will continue to discuss the distinct dynamical behaviors observed when there are two BICs and one BIC, respectively.

\subsection{Two BICs: Rabi oscillation}
\label{BIC1}

In this subsection, we focus on the case of $n_{1}=1, n_{2}=7, m_{1}=4, m_{2}=10$, which implies $N_1 = N_2 = 6$ and $\Delta = 3$. In this scenario, the system is characterized by a pair of BICs, symmetrically located above and below the resonant frequency of the atoms and the bare resonators under the condition $\Omega_1 = \Omega_2 = \omega_c$, as shown in Table~I.

\begin{figure}
\centering
\includegraphics[width=1\columnwidth]{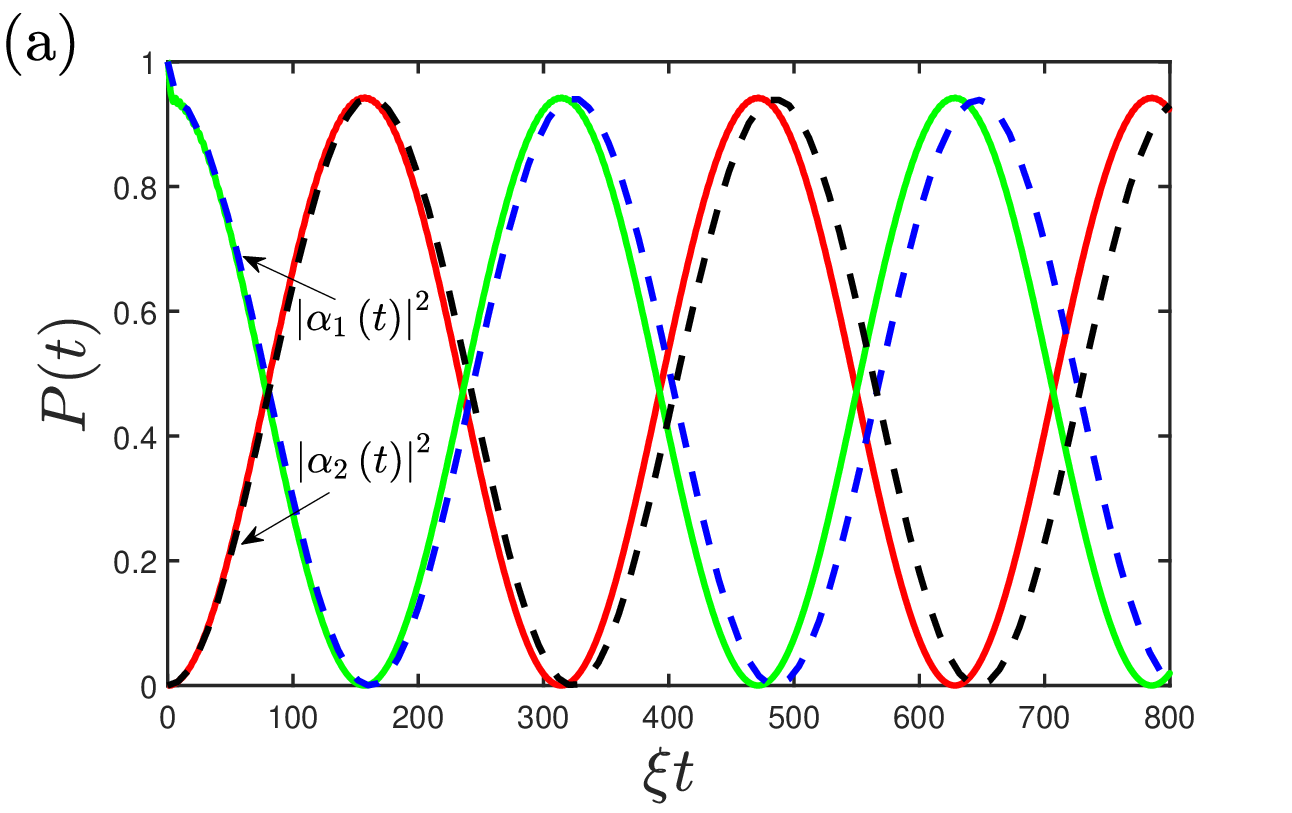}
\includegraphics[width=1\columnwidth]{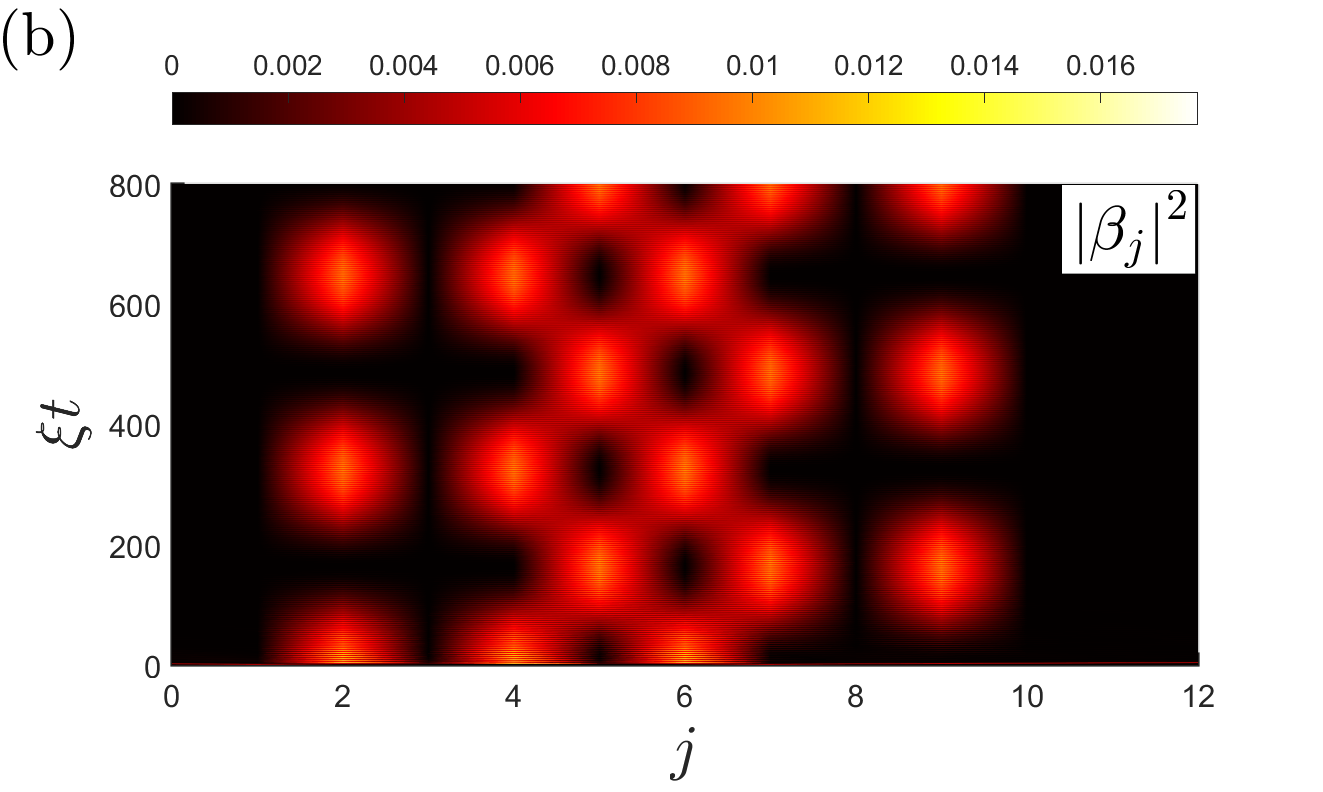}
\includegraphics[width=1\columnwidth]{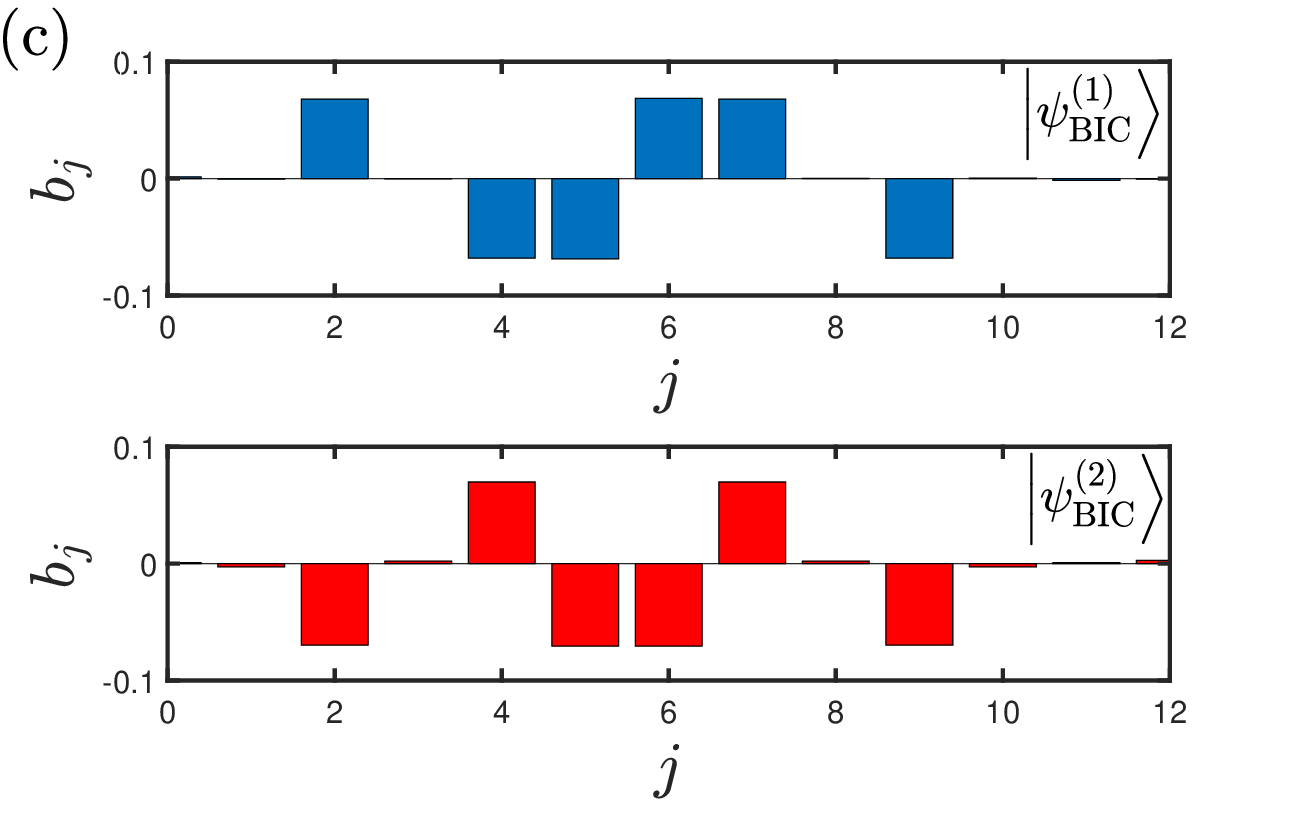}
\includegraphics[width=1\columnwidth]{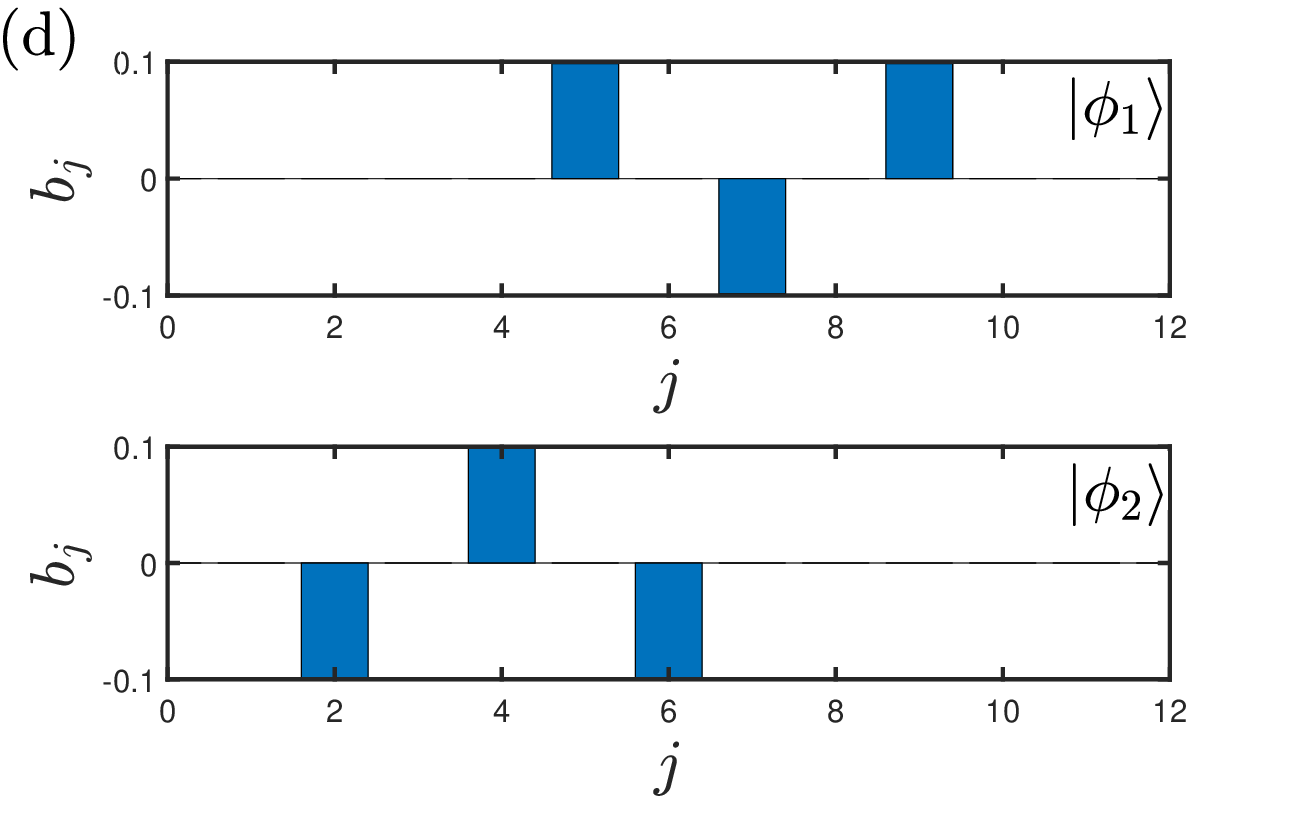}
\caption{(a) The population dynamics of the two giant atoms.
(b) The dynamic evolution of the photonic distribution in real space.
(c) The photonic distributions of the BICs.
(d) The photonic distributions of two separate BICs.
The parameters are set as $\Omega_{1} = \Omega_{2} = \omega_{c} = 0$, $g_{1} = g_{2} = 0.1\xi$, $n_1 = 1$, $n_2 = 7$, $m_1 = 4$, and $m_2 = 10$.}
\label{oscillation}
\end{figure}

Starting from the initial state $|\psi(0)\rangle = \sigma_1^+|G\rangle$, we plot the dynamics of the atomic populations $|\alpha_{1}(t)|^{2}$ and $|\alpha_{2}(t)|^{2}$ in Fig.~\ref{oscillation}(a). The solid lines represent the analytical results derived from Eq.~(\ref{dyn1}), while the dashed lines correspond to the numerical results obtained from $|\psi(t)\rangle = \exp(-iHt)|\psi(0)\rangle$. The excellent agreement between these two results confirms the validity of our Weisskopf-Wigner approximation. We observe that the excitation undergoes a Rabi oscillation between the two giant atoms, with the oscillation period $T = 2\pi / (E_1 - E_2)$ determined by the energy difference between the two BICs.

A similar oscillation is observed in the photonic dynamical evolution, as demonstrated in Fig.~\ref{oscillation}(b). Initially, the photon emitted by the first giant atom is confined within the region it covers, specifically at $j = n_1 + 1, n_1 + 3$, and $n_1 + 5$. As the excitation exchange between the two giant atoms completes, the photon in the waveguide is transferred to the region covered by the second giant atom, at $j = m_1 + 1, m_1 + 3$, and $m_1 + 5$. Over time, the photon periodically oscillates between these two regions, with the same period as the atomic oscillation shown in Fig.~\ref{oscillation}(a).

To further explore the physical mechanism underlying the observed Rabi oscillation, we numerically plot the photonic distributions of the two BICs ($|\psi_{\rm BIC}^{(1,2)}\rangle$) in Fig.~\ref{oscillation}(c), where $b_j$ represents the photon amplitude at the $j$th site of the waveguide. By comparing Figs.~\ref{oscillation}(b) and (c), we observe that the system oscillates between the states:
\begin{eqnarray}
|\phi_1\rangle \sim |\psi_{\rm BIC}^{(1)}\rangle - |\psi_{\rm BIC}^{(2)}\rangle, \quad
|\phi_2\rangle \sim |\psi_{\rm BIC}^{(1)}\rangle + |\psi_{\rm BIC}^{(2)}\rangle.
\end{eqnarray}

Recalling that a single BIC also exists in the cases of $g_1 = 0,g_2\neq 0$ and $g_1\neq0, g_2 = 0$, respectively, the photonic distributions of these two BICs are shown in Fig.~\ref{oscillation}(d). The results indicate that these two BICs correspond to $|\phi_1\rangle$ and $|\phi_2\rangle$. Thus, the Rabi oscillation observed in this subsection arises from the interaction between the BICs formed individually by the first and second giant atoms. This demonstrates that the waveguide acts as a data bus, mediating an effective coupling between the two giant atoms.

\subsection{One BIC: Fractional dissipation}

\begin{figure}
\centering
\includegraphics[width=1\columnwidth]{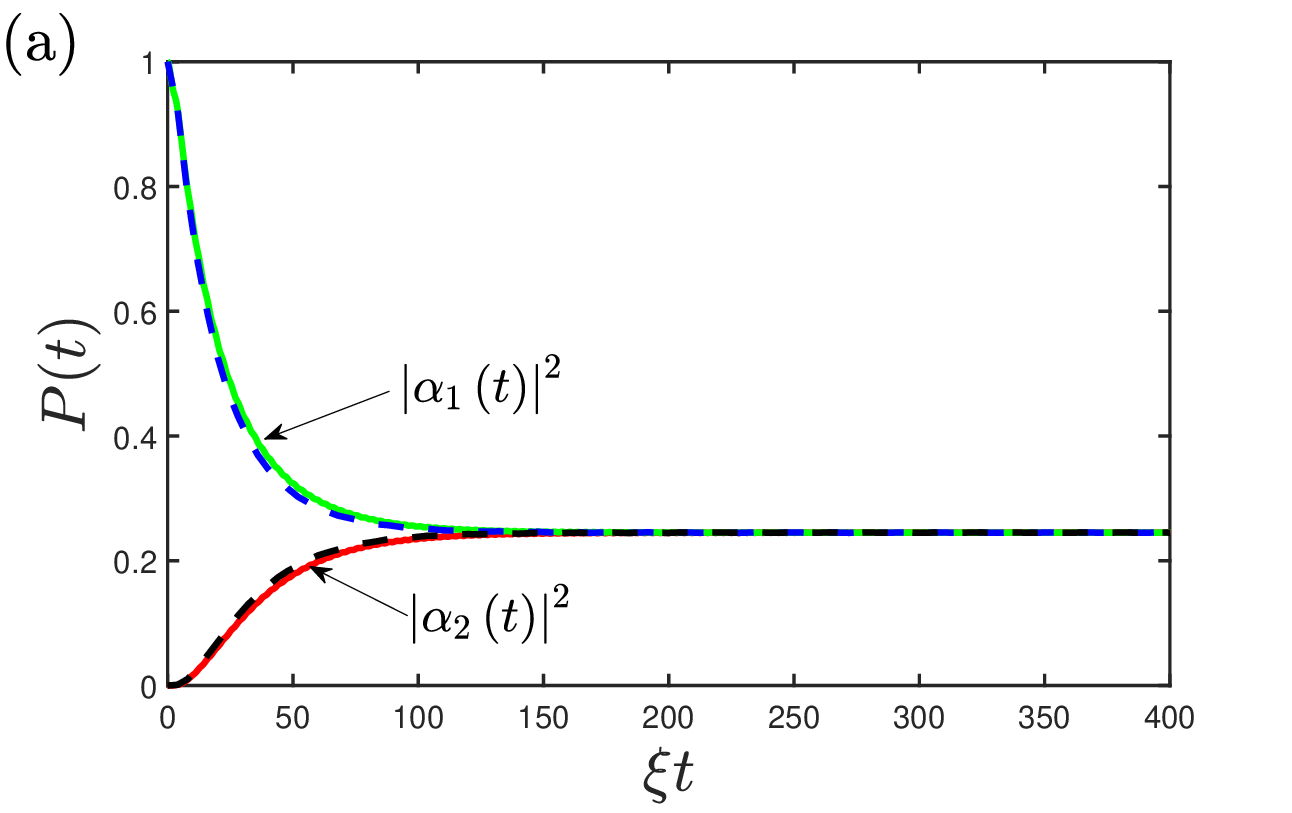}
\includegraphics[width=1\columnwidth]{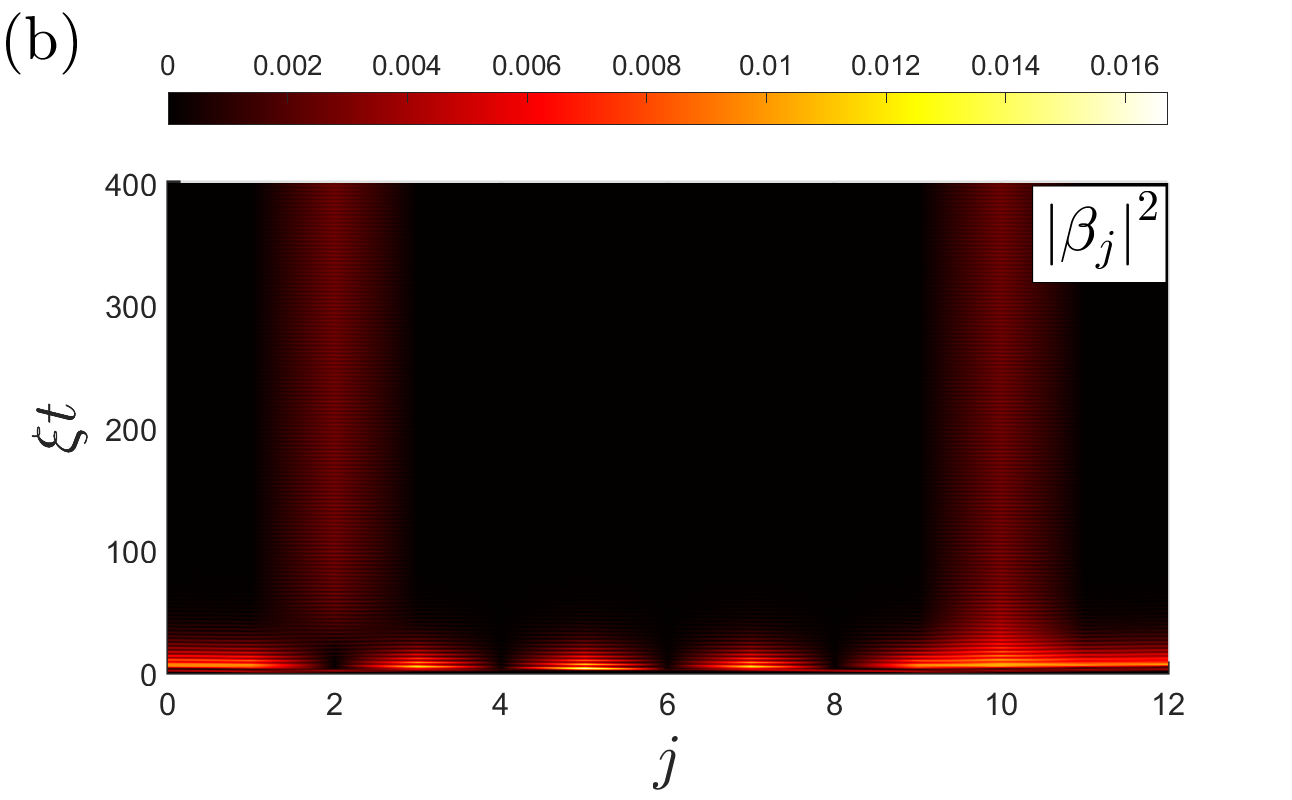}
\includegraphics[width=1\columnwidth]{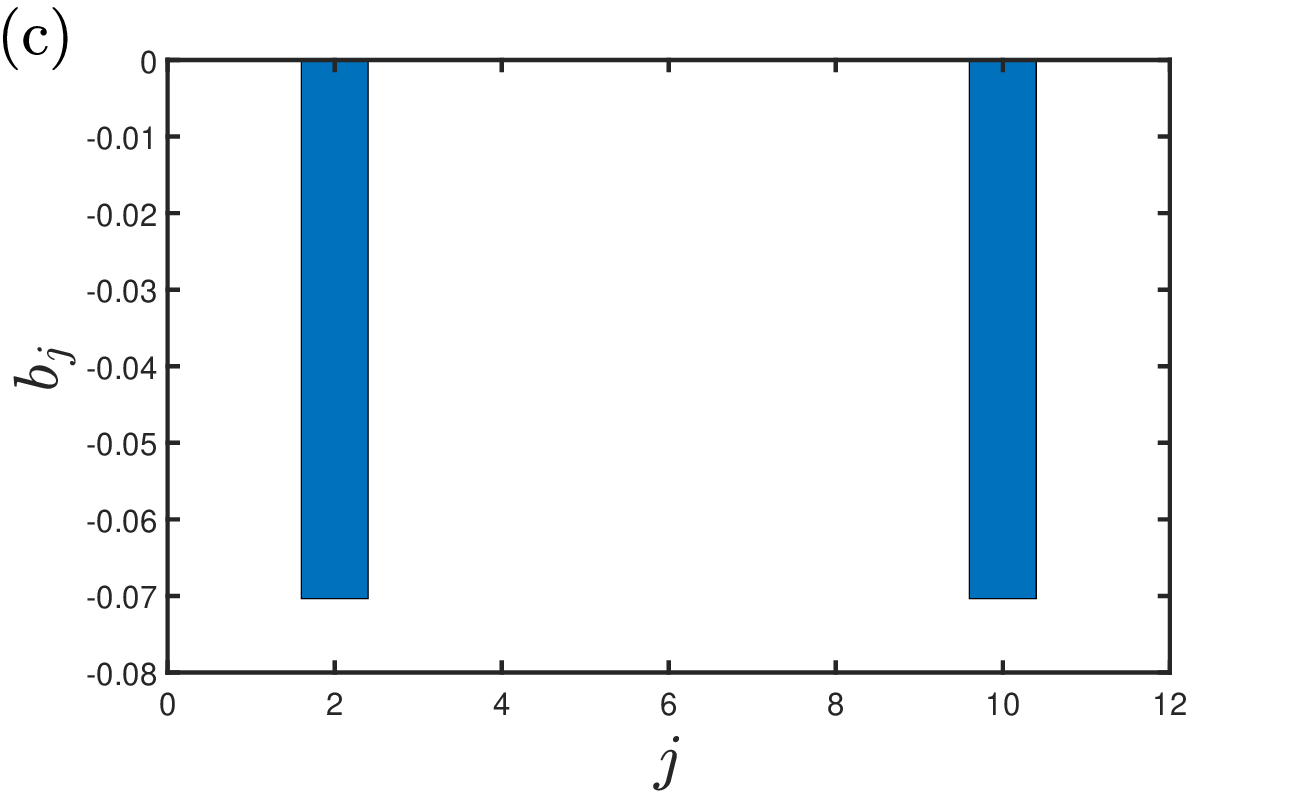}
\caption{(a) The population dynamics of the two giant atoms.
(b) The dynamic evolution of the photonic distribution in real space.
(c) The photonic distribution of the BIC.
The parameters are set as $\Omega_{1} = \Omega_{2} = \omega_{c} = 0$, $g_{1} = g_{2} = 0.1\xi$, $n_1 = 1$, $n_2 = 9$, $m_1 = 3$, and $m_2 = 11$.}
\label{incomplete}
\end{figure}

Now, we consider the case of $n_1=1, n_2=9, m_1=3, m_2=11$, corresponding to $N_1 = N_2 = 8$ and $\Delta = 2$. As shown in Table~I, this configuration supports only one BIC.

Starting with the same initial state $|\psi(0)\rangle = \sigma_1^+|G\rangle$ as in the previous subsection, the atomic populations as functions of evolution time are plotted in Fig.~\ref{incomplete}(a), using both numerical (solid lines) and analytical (dashed lines) approaches. The dynamics in this case is characterized by the fractional population of the first giant atom, with both giant atoms eventually reaching the same steady state as the evolution time approaches infinity. Here, the term ``fractional population"~\cite{JT2022} indicates that the giant atom does not decay completely to the ground state, despite undergoing dissipation due to its coupling to the waveguide.

The photonic dynamics is further analyzed in Fig.~\ref{incomplete}(b), which shows that the photon emitted by the first giant atom initially spreads throughout the waveguide, both inside and outside the regions covered by the giant atoms. As time progresses, the photon gradually diffuses along the waveguide, and the residual photons eventually occupy the $2$nd and $10$th sites without further spreading. This final steady state corresponds to the system's BIC, whose photonic distribution is presented in Fig.~\ref{incomplete}(c). The photonic amplitude exhibits a symmetric profile.

The consistency between the steady state and the BIC is further verified mathematically. Numerically, we find that the atomic excitation probabilities satisfy:
\begin{eqnarray}
\label{P2}
|\alpha_1(t \rightarrow \infty)|^2 &=& |\alpha_2(t \rightarrow \infty)|^2 \nonumber \\
&=& |\langle \psi(0)|E_{\rm BIC}\rangle\langle E_{\rm BIC}|\psi(0)\rangle|^2.
\end{eqnarray}
This indicates that the initial state primarily projects onto the BIC and the extended states. Over time, the components in the extended states vanish due to interference effects, leaving only the BIC component. Consequently, the system evolves into the final steady state.

\section{Conclusion}
\label{consection}

In conclusion, we have studied the dynamics of two giant atoms coupled to a CRW in a braided configuration. Beyond the pioneering experimental work on the coupling between transmons and surface acoustic waves~\cite{MV2014}, giant atoms have since been coupled to waveguides at more than two distinct locations~\cite{BK2020}. Recently, the topology of two nested giant spin ensembles (GSEs) has been demonstrated~\cite{ZW2022}, providing a novel platform for studying ``giant atom'' physics. Moreover, CRWs constructed with nine superconducting qubits have been realized experimentally, where the nearest-neighbor coupling strength is $\xi/(2\pi) = 50$ MHz~\cite{PR2017}. In a recent development, Painter's group constructed a CRW consisting of a $42$-unit-cell array of microwave resonators~\cite{XZ2023}, achieving a qubit-resonator coupling strength of $g/(2\pi) = 124.6$ MHz~\cite{EK2021}.

We have consistently connected the dynamical behaviors of the system to the presence of BICs. When the system supports two BICs, we observe nearly perfect Rabi oscillations, indicating effective atomic interactions mediated by the waveguide. For the case of a single BIC, we predict fractional population dynamics as the system evolves to its final steady state.

These findings demonstrate that the environment does not always lead to complete dissipation in an open quantum system. With the assistance of bound states, particularly the BICs, it is possible to protect the quantum system from dissipation into surrounding reservoirs. This highlights the potential of utilizing waveguide QED to advance quantum technologies.

\begin{acknowledgments}
Z. W. is supported by the Science and Technology Development Project of Jilin Province (Grant No. 20230101357JC) and National Science Foundation of China (Grant No. 12375010).
\end{acknowledgments}

\appendix
\addcontentsline{toc}{section}{Appendices}\markboth{APPENDICES}{}

\begin{subappendices}
\begin{widetext}

\section{THE ENERGY EIGENVALUES OF THE SYSTEM}
\label{A2}
In this appendix, we provide the derivation of Eq.~(\ref{EBIC}) from the main text.

We consider the resonance condition $\Omega_1 = \Omega_2 = \Omega$ and $g_1 = g_2 = g$. {Due to the spatial reflection symmetry of the environment, the modes $\pm k$ are degenerate in $\omega_{k}$.} Using this symmetry and eliminating $A_1$ and $A_2$ in Eqs.~(\ref{a11}) and (\ref{a22}) of the main text, the eigenenergy of the system satisfies:
\begin{eqnarray}
E &=& \Omega + \frac{g^{2}}{N_{c}} \sum_{k} \frac{2 + \cos(kN) \pm \sum_{j,j'=1}^{2} \cos[k(m_{j} - n_{j'})]}{E - \omega_{k}} \nonumber \\
&=& \Omega + \frac{g^{2}}{2\pi} \intop_{-\pi}^{\pi} \frac{2 + 2\cos(kN) \pm \sum_{j,j'=1}^{2} \cos[k(n_{j} - m_{j'})]}{E - \omega_{c} - 2\xi\cos k} \, dk,
\end{eqnarray}
where the dispersion relation is $\omega_{k} = \omega_{c} - 2\xi\cos k$.

To simplify the calculation, let $z = e^{ik}$ and use the integral relation $\ointctrclockwise_{z=1} dz = \ointclockwise_{z^{-1}=1} dz^{-1}$. The above equation becomes:
\begin{eqnarray}
E & = & \Omega+\frac{g^{2}}{2\pi i}\ointctrclockwise_{z=1}\frac{2+z^{N}+z^{-N}\pm\frac{1}{2}\sum_{j,j'=1}^{2}\left(z^{n_{j}-m_{j'}}+z^{-\left(n_{j}-m_{j'}\right)}\right)}{z\left(E-\omega_{c}\right)-\xi\left(z^{2}+1\right)}dz\nonumber \\
 & = & \Omega-\frac{g^{2}}{\xi}\ointctrclockwise_{z=1}\frac{1+z^{N}}{\left(z-\chi-i\epsilon\right)\left(z-\chi^{*}-i\epsilon\right)}\frac{dz}{2\pi i}+\frac{g^{2}}{\xi}\ointclockwiseop_{z'=1}\frac{1+z'^{N}}{\left[z'-\left(\chi+i\epsilon\right)^{-1}\right]\left[z'-\left(\chi^{*}+i\epsilon\right)^{-1}\right]}\frac{dz'}{2\pi i}\nonumber \\
 &  & \pm\frac{g^{2}}{2\xi}\ointctrclockwise_{z=1}\frac{\sum_{j,j'=1}^{2}z^{n_{j}-m_{j'}}}{\left(z-\chi-i\epsilon\right)\left(z-\chi^{*}-i\epsilon\right)}\frac{dz}{2\pi i}\pm\frac{g^{2}}{2\xi}\ointclockwiseop_{z'=1}\frac{\sum_{j,j'=1}^{2}z'^{\left(n_{j}-m_{j'}\right)}}{\left[z'-\left(\chi+i\epsilon\right)^{-1}\right]\left[z'-\left(\chi^{*}+i\epsilon\right)^{-1}\right]}\frac{dz'}{2\pi i},
\end{eqnarray}
where $\chi$ and $\chi^{*}$ are the solutions to the equation $z^{2} - z(E - \omega_{c})/\xi + 1 = 0$, and $\epsilon$ is an infinitesimally small positive value. The specific form of $\chi$ is:
\begin{equation}
\chi = \frac{E - \omega_{c}}{2\xi} - i \sqrt{1 - \left(\frac{E - \omega_{c}}{2\xi}\right)^{2}} = \exp\left[-i\arccos\left(\frac{E - \omega_{c}}{2\xi}\right)\right],
\end{equation}
which satisfies $\chi\chi^{*} = 1$.

Using the residue theorem, each integral in the equation can be evaluated, yielding:
\begin{eqnarray}
E &=& \Omega - \frac{g^{2}}{\xi} \left[\frac{1 + \chi^{N}}{\chi - \chi^{*}} + \frac{1 + (\chi^{*})^{-N}}{(\chi^{*})^{-1} - \chi^{-1}}\right] \nonumber \\
& & \pm \frac{g^{2}}{2\xi} \sum_{j,j'=1}^{2} \left[\frac{\chi^{n_{j} - m_{j'}}}{\chi - \chi^{*}} + \frac{(\chi^{*})^{-(n_{j} - m_{j'})}}{(\chi^{*})^{-1} - \chi^{-1}}\right] \nonumber \\
&=& \Omega - \frac{2g^{2}}{\xi} \frac{1 + \chi^{N}}{\chi - \chi^{*}} \pm \frac{g^{2}}{\xi} \sum_{j,j'=1}^{2} \frac{\chi^{n_{j} - m_{j'}}}{\chi - \chi^{*}} \nonumber \\
&=& \Omega - \frac{g^{2}}{\xi} \frac{2 + 2\chi^{N} \pm \sum_{j,j'=1}^{2} \chi^{n_{j} - m_{j'}}}{\chi - \chi^{*}}.
\end{eqnarray}
This completes the derivation of Eq.~(\ref{EBIC}).

\section{NON-MARKOVIAN DYNAMICS}
\label{A1}
In this appendix, we provide the detailed derivation of the dynamical equation presented in the main text.

By introducing the Fourier transform $a_{j} = \frac{1}{\sqrt{N_{c}}} \sum_{k} e^{-ikj} b_{k}$, the total Hamiltonian in the momentum space is written as:
\begin{eqnarray}
\label{toalhamition}
H &=& \Omega_{1}\left|e\right\rangle_{1}\left\langle e\right| + \Omega_{2}\left|e\right\rangle_{2}\left\langle e\right| + \sum_{k}\omega_{k}b_{k}^{\dagger}b_{k} \nonumber \\
& & + \frac{g_{1}}{\sqrt{N_{c}}} \sum_{k} \left[\left(e^{ikn_{1}} + e^{ikn_{2}}\right) b_{k}^{\dagger} \sigma_{1}^{-} + {\rm H.c.}\right] \nonumber \\
& & + \frac{g_{2}}{\sqrt{N_{c}}} \sum_{k} \left[\left(e^{ikm_{1}} + e^{ikm_{2}}\right) b_{k}^{\dagger} \sigma_{2}^{-} + {\rm H.c.}\right].
\end{eqnarray}
In the single excitation subspace, the time-dependent wave function is written as:
\begin{equation}
\label{wavet}
\left|\psi\left(t\right)\right\rangle = \left[\alpha_{1}\left(t\right) \sigma_1^+ + \alpha_{2}\left(t\right) \sigma_2^+ + \sum_{k} \beta_{k}\left(t\right) b_k^{\dagger}\right] \left|G\right\rangle.
\end{equation}
From the Schr\"{o}dinger equation, we obtain the following three equations:
\begin{eqnarray}
\label{alpha1}
i\frac{\partial}{\partial t}\alpha_{1}\left(t\right) &=& \Omega_{1}\alpha_{1}\left(t\right) + \frac{g_{1}}{\sqrt{N_{c}}} \sum_{k} \left(e^{-ikn_{1}} + e^{-ikn_{2}}\right) \beta_{k}\left(t\right), \\
i\frac{\partial}{\partial t}\alpha_{2}\left(t\right) &=& \Omega_{2}\alpha_{2}\left(t\right) + \frac{g_{2}}{\sqrt{N_{c}}} \sum_{k} \left(e^{-ikm_{1}} + e^{-ikm_{2}}\right) \beta_{k}\left(t\right), \\
\label{beltak}
i\frac{\partial}{\partial t}\beta_{k}\left(t\right) &=& \omega_{k}\beta_{k}\left(t\right) + \frac{g_{1}}{\sqrt{N_{c}}} \left(e^{ikn_{1}} + e^{ikn_{2}}\right) \alpha_{1}\left(t\right) \nonumber \\
& & + \frac{g_{2}}{\sqrt{N_{c}}} \left(e^{ikm_{1}} + e^{ikm_{2}}\right) \alpha_{2}\left(t\right).
\end{eqnarray}
Assuming that all resonators in the waveguide are initially in their vacuum states, i.e., $\beta_{k}\left(0\right) = 0$, we find:
\begin{eqnarray}
\beta_{k}\left(t\right) &=& -\frac{ig_{1}}{\sqrt{N_{c}}} \left(e^{ikn_{1}} + e^{ikn_{2}}\right) \int_{0}^{t} d\tau \alpha_{1}\left(\tau\right) e^{-i\omega_{k}\left(t-\tau\right)} \nonumber \\
& & -\frac{ig_{2}}{\sqrt{N_{c}}} \left(e^{ikm_{1}} + e^{ikm_{2}}\right) \int_{0}^{t} d\tau \alpha_{2}\left(\tau\right) e^{-i\omega_{k}\left(t-\tau\right)}.
\end{eqnarray}
Substituting this expression into Eq.~(\ref{alpha1}) and integrating over $k$, we obtain:
\begin{eqnarray}
\frac{\partial}{\partial t}\alpha_{1}\left(t\right) &=& -i\Omega_{1}\alpha_{1}\left(t\right) - 2g_{1}^{2} \int_{0}^{t} d\tau \alpha_{1}\left(\tau\right) e^{-i\omega_{c}\tau} \left\{ J_{0}\left(2\xi\tau\right) + i^{N_{1}} J_{N_{1}}\left(2\xi\tau\right) \right\} \nonumber \\
& & - g_{1}g_{2} \int_{0}^{t} d\tau \alpha_{2}\left(\tau\right) e^{-i\omega_{c}\tau} \sum_{j,j'=1}^{2} i^{n_{j} - m_{j'}} J_{n_{j} - m_{j'}}\left(2\xi\tau\right).
\end{eqnarray}
Following a similar procedure, the equation for $\alpha_{2}(t)$ is:
\begin{eqnarray}
\frac{\partial}{\partial t}\alpha_{2}\left(t\right) &=& -i\Omega_{2}\alpha_{2}\left(t\right) - 2g_{2}^{2} \int_{0}^{t} d\tau \alpha_{2}\left(\tau\right) e^{-i\omega_{c}\tau} \left\{ J_{0}\left(2\xi\tau\right) + i^{N_{2}} J_{N_{2}}\left(2\xi\tau\right) \right\} \nonumber \\
& & - g_{1}g_{2} \int_{0}^{t} d\tau \alpha_{1}\left(\tau\right) e^{-i\omega_{c}\tau} \sum_{j,j'=1}^{2} i^{n_{j} - m_{j'}} J_{n_{j} - m_{j'}}\left(2\xi\tau\right).
\end{eqnarray}

Finally, to calculate the photon distribution in real space, we use the inverse Fourier transform $\beta_{j} = \frac{1}{\sqrt{N_{c}}} \sum_{k} \beta_{k} e^{-ikj}$:
\begin{eqnarray}
\beta_{j}(t) &=& -ig_{1} \int_{0}^{t} d\tau \alpha_{1}\left(t-\tau\right) F_{j}\left(\tau\right) \nonumber \\
& & -ig_{2} \int_{0}^{t} d\tau \alpha_{2}\left(t-\tau\right) G_{j}\left(\tau\right),
\end{eqnarray}
where
\begin{eqnarray}
F_{j}\left(\tau\right) &=& e^{-i\omega_{c}\tau} \sum_{q=1}^{2} i^{\left(j-n_{q}\right)} J_{\left(j-n_{q}\right)}(2\xi\tau), \\
G_{j}\left(\tau\right) &=& e^{-i\omega_{c}\tau} \sum_{q=1}^{2} i^{\left(j-m_{q}\right)} J_{\left(j-m_{q}\right)}(2\xi\tau).
\end{eqnarray}
This completes the derivation of the dynamical equations and the photon distribution in real space.


\end{widetext}
\end{subappendices}



\begin{thebibliography}{99}

\bibitem{AD2015} A. D. Ludlow, M. M. Boyd, J. Ye, E. Peik, and P. O. Schmidt, Optical atomic clocks,
                   Rev. Mod. Phys. {\bf 87}, 637 (2015).
\bibitem{CD2019}C. D. Bruzewicz, J. Chiaverini, R. McConnell, and J. M. Sage, Trapped-ion quantum computing: progress and challenges,
                   Appl. Phys. Rev. {\bf 6}, 021314 (2019).
\bibitem{DL2003}D. Leibfried, R. Blatt, C. Monroe, and D. Wineland, Quantum dynamics of single trapped ions, Rev. Mod. Phys. {\bf75}, 281 (2003).
\bibitem{MV2014}M. V. Gustafsson, T. Aref, A. F. Kockum, M. K. Ekstr\"{o}m, G. Johansson, and P. Delsing, Propagating phonons coupled to an
                   artificial atom, Science {\bf346}, 207 (2014).
\bibitem{JK2007}J. Koch, T. M. Yu, J. Gambetta, A. A. Houck, D. I. Schuster, J. Majer, A. Blais, M. H. Devoret, S. M. Girvin, and R. J. Schoelkopf,
                Charge-insensitive qubit design derived from the Cooper pair box, Phys. Rev. A {\bf76}, 042319 (2007).
\bibitem{SD1986}S. Datta, {\it Surface Acoustic Wave Devices} (Prentice Hall, 1986).
\bibitem{DM2007}D. Morgan, {\it Surface Acoustic Wave Filters}, 2nd ed. (Academic Press, 2007).
\bibitem{AF2018} A. F. Kockum, G. Johansson, and F. Nori, Decoherence-free
                        interaction between giant atoms in waveguide quantum electrodynamics, Phys. Rev. Lett. {\bf 120}, 140404 (2018).
\bibitem{BK2020}B. Kannan, M. J. Ruckriegel, D. L. Campbell, A. F. Kockum, J. Braum\"{u}ller, D. K. Kim, M. Kjaergaard, P. Krantz, A. Melville,
                        B. M. Niedzielski, A. Veps\"{a}l\"{a}inen, R. Winik, J. L. Yoder, F.
                        Nori, T. P. Orlando, S. Gustavsson, and W. D. Oliver, Waveguide quantum electrodynamics with superconducting artificial
                         giant atoms, Nature (London) {\bf583}, 775 (2020).
\bibitem{GA2019}G. Andersson, B. Suri, L. Guo, T. Aref, and P. Delsing, Non-exponential decay of a giant artificial atom, Nature Phys, {\bf15}, 1123 (2019).
\bibitem{AM2021}A. M. Vadiraj, A. Ask, T. G. McConkey, I. Nsanzineza, C. W. Sandbo Chang, A. F. Kockum, and C. M. Wilson, Engineering the level structure of a giant artificial atom in waveguide quantum electrodynamics,
    Phys. Rev. A {\bf103}, 023710 (2021).
\bibitem{ZW2022}Z. Wang, Y. Wang, J. Yao, R. Shen, W. Wu, J. Qian, J. Li, S. Zhu, and J. You, Giant spin ensembles in waveguide magnonics,
Nature Commun. {\bf13}, 7580 (2022).
\bibitem{DF2008}D. F. Walls and G. J. Milburn, {\it Quantum Optics}, 2nd ed. (Springer, 2008).
\bibitem{AF2021} A. F. Kockum, Quantum optics with giant atoms-the
                 first five years, {\it International Symposium on Mathematics,
                    Quantum Theory, and Cryptography}, edited by T. Takagi, M. Wakayama, K. Tanaka, N. Kunihiro, K. Kimoto, and Y.
                     Ikematsu (Springer Singapore, Singapore, 2021), p. 125.
\bibitem{WZ2020} W. Zhao and Z. Wang, Single-photon scattering and bound
                 states in an atom-waveguide system with two or multiple coupling points, Phys. Rev. A {\bf101}, 053855 (2020).


\bibitem{XG2017}X. Gu, A. F. Kockum, A. Miranowicz, Y.-X. Liu, and F. Nori, Microwave photonics with superconducting quantum circuits, Phys. Rep.
                     {\bf718}, 1 (2017).
\bibitem{DR2017}D. Roy, C. M. Wilson, and O. Firstenberg, Colloquium: strongly interacting photons in one-dimensional continuum,
            Rev. Mod. Phys. {\bf89}, 021001 (2017).
\bibitem{RH19701} R. H. Lehmberg, Radiation from an N-atom system. I. General formalism, Phys. Rev. A {\bf2}, 883 (1970).
\bibitem{RH19702}R. H. Lehmberg, Radiation from an N-atom system. II. Spontaneous emission from a pair of atoms, Phys. Rev. A {\bf2}, 889 (1970).
\bibitem{KL2013} K. Lalumi\`{e}re, B. C. Sanders, A. F. van Loo, A. Fedorov, A. Wallraff, and A. Blais, Input-output theory for waveguide QED with an ensemble
                    of inhomogeneous atoms, Phys. Rev. A {\bf88}, 043806 (2013).
\bibitem{HZ2013} H. Zheng and H. U. Baranger, Persistent quantum beats and long-distance entanglement from waveguide-mediated interactions, Phys. Rev.
                  Lett. {\bf110}, 113601 (2013).
\bibitem{AC2020}A. Carollo, D. Cilluffo, and F. Ciccarello, Mechanism of decoherence-free coupling between giant atoms,
                   Phys. Rev. Res. {\bf 2}, 043184 (2020).

\bibitem{LG2017}L. Guo, A. Grimsmo, A. F. Kockum, M. Pletyukhov, and
                       G. Johansson, Giant acoustic atom: a single quantum system with a deterministic time delay, Phys. Rev. A {\bf95}, 053821
                       (2017).



\bibitem{CA2021}C. A. Gonz\'{a}lez-Guti\'{e}rrez, J. Rom\'{a}n-Roche, and D. Zueco,
                          Distant emitters in ultrastrong waveguide QED: ground-state
                        properties and non-Markovian dynamics, Phys. Rev. A {\bf104}, 053701 (2021).
\bibitem{LG2020}L. Guo, A. F. Kockum, F. Marquardt, and G. Johansson, Oscillating bound states for a giant atom, Phys. Rev. Res. {\bf2}, 043014
                         (2020).
\bibitem{SG2020}S. Guo, Y. Wang, T. Purdy, and J. Taylor, Beyond spontaneous
                  emission: Giant atom bounded in the continuum, Phys. Rev. A {\bf102}, 033706 (2020).
\bibitem{XL2022}X. Yin, W. Luo, and J. Liao, Non-Markovian disentanglement dynamics in double-giant-atom waveguide-QED
                     systems, Phys. Rev. A {\bf106}, 063703 (2022).
\bibitem{LD2023}L. Du, Y. Zhang, and Y. Li, A giant atom with
                     modulated transition frequency, Front. Phys. {\bf18}, 12301 (2023).
\bibitem{SLF2021}S. L. Feng and W. Z. Jia, Manipulating single-photon transport in a waveguide-QED structure containing two giant atoms,
                        Phys. Rev. A {\bf104}, 063712 (2021).
\bibitem{YC2022}Y. Chen, L. Du, L. Guo, Z. Wang, Y. Zhang, Y. Li, and J.
                Wu, Nonreciprocal and chiral single-photon scattering for giant
                         atoms, Commun. Phys. {\bf5}, 215 (2022).
\bibitem{YL2017}Y. Liu and A. A. Houck, Quantum electrodynamics near a
                          photonic bandgap, Nat. Phys. {\bf13}, 48 (2017).
\bibitem{NM2019}N. M. Sundaresan, R. Lundgren, G. Zhu, A. V. Gorshkov,
                              and A. A. Houck, Interacting qubit-photon bound states with
                           superconducting circuits, Phys. Rev. X {\bf9}, 011021 (2019).
\bibitem{PM2019}P. M. Harrington, M. Naghiloo, D. Tan, and K. W. Murch,
                          Bath engineering of a fluorescing artificial atom with a photonic
                        crystal, Phys. Rev. A {\bf99}, 052126 (2019).

\bibitem{FH1975} F. H. Stillinger and D. R. Herrick, Bound states in the continuum,
                 Phys. Rev. A {\bf11}, 446 (1975).
\bibitem{DC2008}D. C. Marinica, A. G. Borisov, and S. V. Shabanov, Bound
                   states in the continuum in photonics, Phys. Rev. Lett. {\bf100}, 183902 (2008).
\bibitem{MI2012}M. I. Molina, A. E. Miroshnichenko, and Y. S. Kivshar, Surface
                 bound states in the continuum, Phys. Rev. Lett. {\bf108}, 070401 (2012).
\bibitem{GC2019}G. Calaj\'{o}, Y. L. Fang, H. U. Baranger, and F. Ciccarello, Exciting a bound state in the continuum through multiphoton
                      scattering plus delayed quantum feedback, Phys. Rev. Lett. {\bf122}, 073601 (2019).
\bibitem{QQ2023}Q. Qiu, Y. Wu, and X. L\"{u}, Collective radiance of giant atoms
                        in non-Markovian regime, Sci. China Phys. Mech. Astron. {\bf66}, 224212 (2023).

\bibitem{wang2015} Z. Wang, Y. J. Ji, Yong Li, and D. L. Zhou, Dissipation and decoherence induced by collective dephasing in a coupled-qubit
system with a common bath, Phys. Rev. A {\bf 91}, 013838 (2015).

\bibitem{NS2018}N. Shammah, S. Ahmed, N. Lambert, S. De Liberato, and F. Nori, Open quantum systems with local and collective incoherent processes: Efficient numerical simulations using permutational invariance, Phys. Rev. A {\bf 98}, 063815 (2018).
\bibitem{SL2021}S. Longhi, Rabi oscillations of bound states in the continuum,
                  Opt. Lett. {\bf 46}, 2091 (2021).
\bibitem{KH2023}K. H. Lim, W. -K. Mok,  and L. -C. Kwek, Oscillating bound states
                 in non-Markovian photonic lattices, Phys. Rev. A {\bf 107}, 023716 (2023).

\bibitem{CW2016}C. W. Hsu, B. Zhen, A. D. Stone, J. D. Joannopoulos, and M. Soljacic, Bound states in the continuum, Nat. Rev. Mater. {\bf1}, 16048 (2016).
\bibitem{MF2017}M. F. Limonov, M. V. Rybin, A. N. Poddubny, and Y.-S. Kivshar, Fano resonances in photonics, Nature Photon. {\bf11}, 543 (2017).
\bibitem{KK2019}K. Koshelev, A. Bogdanov, and Y. Kivshar, Meta-optics and bound states in the continuum, Sci. Bull. {\bf64}, 836 (2019).
\bibitem{KKK2019}K. Koshelev, G. Favraud, A. Bogdanov, Y. Kivshar, and A. Fratalocchi, Nonradiating photonics with resonant dielectric nanostructures,
                 Nanophoton. {\bf8}, 725 (2019).
\bibitem{JT2022} J. Talukdar and D. Blume, Two emitters coupled to a bath with Kerr-like nonlinearity: Exponential decay, fractional populations, and Rabi oscillations, Phys. Rev. A {\bf105}, 063501 (2022).
\bibitem{ELi1987} E. Yablonovitch, Inhibited spontaneous emission in solid-state physics and electronics, Phys. Rev. Lett. {\bf58}, 2059 (1987) .
\bibitem{Wang1990}S. John and J. Wang, Quantum electrodynamics near a photonic band gap: Photon bound states and dressed atoms,  Phys. Rev. Lett. {\bf64}, 2418 (1990).
\bibitem{An2010}Q. Tong, J. An, H. Luo, and C. H. Oh, Mechanism of entanglement preservation, Phys. Rev. A {\bf81}, 052330 (2010).
\bibitem{An2013}H. Liu, J. An, C. Chen, Q. Tong, H. Luo, and C. H. Oh, Anomalous decoherence in a dissipative two-level system, Phys. Rev. A {\bf87}, 052139 (2013).

\bibitem{SL2008}S. Longhi, Optical bloch oscillations and zener tunneling with nonclassical light, Phys. Rev. Lett. {\bf101}, 193902 (2008).
\bibitem{SL2020}S. Longhi, Photonic simulation of giant atom decay, Opt. Lett. {\bf45}, 3017 (2020).
\bibitem{Zhou2008} L. Zhou, Z. Gong, Y. Liu, C. Sun, and F. Nori, Controllable scattering of a single photon inside a one-dimensional resonator waveguide, Phys. Rev. Lett. {\bf101}, 100501 (2008).
\bibitem{Zhou2013} L. Zhou, L. Yang, Y. Li, and C. Sun, Quantum routing of single photons with a cyclic three-level system, Phys. Rev. Lett. {\bf111}, 103604 (2013).
\bibitem{Yu2021}H. Yu, Z. Wang, and J. Wu, Entanglement preparation and nonreciprocal excitation evolution in giant atoms by controllable  dissipation and coupling, Phys. Rev. A {\bf104}, 013720 (2021).
\bibitem{XJ2023}X.  Zhang, C.  Liu, Z. Gong, and Z. Wang, Quantum interference and controllable magic cavity QED via a giant atom in a coupled resonator waveguide, Phys. Rev. A {\bf 108}, 013704 (2023).

\bibitem{PR2017} P. Roushan, C. Neill, J. Tangpanitanon, V. M. Bastidas,
A. Megrant, R. Barends, Y. Chen, Z. Chen, B. Chiaro, A.
Dunsworth,  et al., Spectroscopic signatures of localization with interacting photons in superconducting qubits,  Science {\bf 358}, 1175 (2017).

\bibitem{XZ2023}X. Zhang, E. Kim, D. K. Mark, S. Chol, and O. Painter, A superconducting quantum simulator based on a photonic-bandgap metamaterial, Science {\bf 379}, 278 (2023).


\bibitem{EK2021}E. Kim, X. Y. Zhang, V. S. Ferreira, J. Banker, J. K. Iverson, A. Sipahigil, M. Bello, A. Gonz\'{a}lez-Tudela, M. Mirhosseini, and
O. Painter, Quantum electrodynamics in a topological waveguide, Phys. Rev. X {\bf 11}, 011015 (2021).







%





\end{thebibliography}
\end{document}